# Nucleosome Switching

David J. Schwab, Robijn F. Bruinsma, Joseph Rudnick

*Department of Physics and Astronomy, University of California, Los Angeles,*

*Los Angeles, CA, 90024*

&

Jonathan Widom

*Department of Biochemistry, Molecular Biology, and Cell Biology, Northwestern*

*University, Evanston. IL, 60208*

**Abstract:**

We present a statistical-mechanical analysis of the positioning of nucleosomes along one of the chromosomes of yeast DNA as a function of the strength of the binding potential and of the chemical potential of the nucleosomes. We find a significant density of two-level nucleosome *switching regions* where, as a function of the chemical potential, the nucleosome distribution undergoes a "micro" first-order transition. The location of these nucleosome switches shows a strong correlation with the location of transcription-factor binding sites.

The chromosomes that contain the genomic code of an organism are densely packed inside the cell nucleus. They are composed of centimeters long DNA molecules, with the *nucleosome* as their basic structural organization unit. Nucleosomes consist of a cluster of eight, positively charged, histone proteins that is tightly wrapped around by a - negatively charged - 147 basepair (bp) stretch of DNA [1]. Between 75 to 90 % of genomic DNA is wrapped around nucleosomes this way, separated by 10-50 bp linkers [2]. In an early model, Kornberg and Stryer [3] (KS) treated the nucleosomes as a *one-dimensional liquid of hard rods* with an excluded volume of the order of 147 bp's. They attributed regularities in nucleosome positioning observed *in-vivo* to the decaying density oscillations near a boundary that, for higher densities, characterize such a liquid [4]. The "boundaries" would be provided here by sequence-specific DNA/protein binding sites of, for example, Transcription Factors (TF's), that would block nucleosome binding. It has long been known that DNA not only codes for the amino-acid sequence of proteins but also for the binding sites of TF's, which play an important regulatory role in gene expression [5].

Recently, Segal et al. [6] provided evidence that the binding sites of nucleosomes are likewise determined by the DNA sequence. The specificity of DNA/nucleosome binding is due here to the sequence-dependent *bending stiffness* of DNA [7]. Through bioinformatics methods, Segal et al. [6] constructed a DNA-nucleosome library of binding probabilities representing all possible nucleotide combinations of a 147 bp DNA sequence. They then treated the logarithm of this binding-probability as an *on-site potential* for a single nucleosome bound to DNA and included this on-site potential in the KS liquid-of-hard-rods model. Segal et al. [6] showed that the thermal equilibrium density profile of this model correctly predicted about *half* of the actual stable nucleosome positions of the yeast genome. A functional role for nucleosome positioning is supported by many studies [8], but how much of nucleosome positioning is a consequence of DNA sequence preference and how much the activity of specific enzymes remains under discussion [9]. Segal et al. [6] did find that the most probable location for TATA elements places them just outside a stably positioned nucleosome, which suggests a possible mechanism for directing transcriptional machinery.

The aim of the present paper is to examine the phase behavior of the model of Segal et al. [6], in terms of temperature and chemical potential, and to interpret the phase behavior in the context of a possible regulatory role for nucleosome positioning. With "temperature" we do not really mean here the actual ambient temperature of the DNA-nucleosome system but rather the overall reference energy scale of the on-site nucleosome/DNA potential (which is not obtained from the bioinformatics analysis). Under *in-vivo* conditions, the energy scale depends on the choice of the DNA sequences, and under *in-vitro* conditions it can be tuned by the ambient salt concentration. The chemical potential is a second key thermodynamic parameter: *in vivo* nucleosome densities vary from 75 to 90 percent between different cell types. If slight changes in average density would alter the nucleosome density profile in a drastic and chaotic manner, then the predicted nucleosome positions would not be "robust" and any functionality for nucleosome positioning would be in serious doubt.

We work in the grand-canonical ensemble with the N-nucleosome Hamiltonian of Segal et al:

$$H = \sum_{i=1}^{N} U(n_{i+1} - n_i) + \sum_{i=1}^{N} V_{n_i} \qquad (1.1)$$

Here, $n_i$ is the location of the first base pair blocked by the ith nucleosome, $U(m)$ is equal to zero if $m$ exceeds the hard-core size $a = 157$ [10] and infinite if $m$ is less than or equal to $a$. Finally, $V_n \equiv -k_B T_0 \log P_n$ with $T_0$ the reference temperature and $P_n$ the binding probability for site $n$ of chromosome II of *Saccharomyces cerevisiae* (budding yeast). The probability distribution of the onsite potentials is approximately Gaussian with a width $\Delta V / k_B T_0$ 9.3. The auto-correlation function $C(i) \propto \langle V_n V_{n+i} \rangle - \langle V_n \rangle^2$ is shown in Fig.1. The decay length of the correlation function is of the order of the size of a nucleosome and the oscillations reflect the 10 bp DNA repeat length.

In the absence of the excluded volume interactions, the nucleosome binding probability $\rho_n$ in the grand-canonical ensemble is just the *Langmuir Isotherm* $\rho_n = \frac{e^{\beta(\mu - V_n)}}{1 + e^{\beta(\mu - V_n)}}$ for the binding of solute molecules to discrete sites, with $\mu$ the chemical potential of the solute molecules. In the presence of excluded-volume interactions, the site potential $V_n$ in the Langmuir Isotherm must be replaced by $V_n - k_b T \ln H_n$ where $H_n = \prod_{m=2}^{a}(1 - h_{n+m-1})$ describes the different ways a site can be blocked by neighboring nucleosomes. Here, $h_i$ are effective site probabilities defined as:

$$h_i = \frac{\rho_i}{1 - \sum_{j=1}^{a-1} \rho_{i-j}} \qquad (1.2)$$

In terms of these effective site probabilities, the binding isotherm:

$$h_n = \frac{H_n e^{\beta(\mu-V_n)}}{1+H_n e^{\beta(\mu-V_n)}} \tag{1.3}$$

adopts the form of a *recursion relation* that can be carried out efficiently by numerical methods. After the $h_i$'s have been found, the actual binding probability profile can be iteratively obtained from Eq. (1.2) and the thermodynamic properties of the system can be obtained from the grand partition function:

$$\Xi(\mu,T) \propto \prod_n \left(\frac{1}{1-h_n}\right) \tag{1.4}$$

The Equation of State, for example, follows from the relation $PL = k_B T \ln \Xi$ (with P the pressure and L the sample length). In the continuum limit, Eq. 1.3 reduces to the density profile of a fluid of hard-rods in an external potential that was obtained by Percus [11].

Figure 2 shows the numerically computed heat capacity at $\mu = 24\ k_B T_0$ (with average site occupancy of about 80%) as a function of the reduced temperature $T/T_0$ for chromosome II of budding yeast. The heat capacity is characterized by a pronounced maximum at a reduced temperature larger than one. On the high temperature side of the maximum, the site occupation probability is approximately *sinusoidal* (right inset, Fig.2). In linear-response theory, an effectively weak external potential produces a density modulation $\rho(q) \propto S(q)V(q)$ that is proportional to the Fourier Transform $V(q)$ of the on-site potential $V_n$. Here, $S(q)$ is the *structure factor* of the hard-rod liquid. The analytical form [11] of $S(q)$ develops, at higher densities, a pronounced maximum around $q = 2\pi/a$. That means that the Fourier component $V(q = 2\pi/a)$ is amplified with respect to the other Fourier components. On the low temperature side, the occupation probability is, for many sites, either close to one or close to zero (left inset, Fig.2) and the linear-response description is clearly invalid. We thus interpret the heat-capacity maximum as a freezing/localization transition. Because this is a one-dimensional system with short-range interactions, it would be expected that nucleosome freezing is not a true phase-transition. Nevertheless, for the special case that the on-site potential $V_n$ is a sine wave with a wavelength equal to the hard-core size $a$ plus a random phase - as is effectively the case in the linear-response regime $T/T_0 \gg 1$ - the freezing has been shown to be a true (continuous) phase transition [12], so this question must remain open. Independently, effective nucleosome positioning clearly requires the system to be on the low-temperature side of the heat capacity maximum.

To clarify the thermodynamics of the low temperature regime, we studied the stability of the density profile with respect to changes in the chemical potential $\mu$ for fixed $T = T_0/2$. We found that, at a given mean occupancy in the biological range (between 75 and 90%), DNA sequences could be divided into two classes in terms of the stability of nucleosome positioning. Most sequences have a well-defined density profile that indeed is robust with respect to changes in the chemical potential. However, a substantial minority of sequences appears to be quite unstable, with poorly defined density profiles and a rapid variation of site probabilities with chemical potential. An example is shown in Fig. 3 where we plot the site occupancy probability of a typical short DNA section of 3,000 bp belonging to this second class for the case that the mean occupancy was 89.3% and $T/T_0=0.5$. Interestingly, when we slightly increased the mean occupancy to 90% the section *did* have a stable, well-defined nucleosome configuration with $P = 11$ nucleosomes while at the slightly lower mean occupancy of 88.8% the section *again* had a stable configuration, now with $P – 1$ nucleosomes. The 'disordered' region for intermediate values of occupancy (or chemical potential) is the *superposition* of the two stable $P$ and $P – 1$ configurations. The "unstable" sections are in actuality localized, *two-level systems* (or multilevel systems with a limited number of competing configurations). The two competing $P$ and $P – 1$ configurations are, of course, not *exactly* degenerate at a given value of the chemical potential. If $\Delta$ is the energy difference between the two levels then, as a function of temperature, the contribution to the heat capacity from a two-level system has a *maximum* when $k_BT$ is of the order of $\Delta$. The heat capacity at a given temperature is dominated by those two-level systems that happen to have an energy difference $\Delta$ of the order of $k_BT$. The low-temperature drop of the heat capacity thus reflects the fact that as the temperature is lowered, it becomes harder and harder to find two level systems that are degenerate on a scale of $k_BT$.

We plotted in Fig.2 the fraction of yeast chromosome II that participates in two-level switching. Whether a sequence classifies as a two-level system is measured here by choosing an arbitrary cutoff (taken to be 0.18) in the density profile. We took a moving window of length 147 bp and counted the number of density spikes that passed through this cutoff. If this number was greater than one, then we marked the region as an unstable section. The fraction of yeast chromosome II genome that participates in switching behavior is shown in Fig. 2 as a function of temperature. The reason for the maximum is that at low temperatures the two-level systems "freeze-out", as noted, while two-level systems disappear at higher temperatures because *most* typical site probabilities drop below the cut-off as a consequence of the thermal melting of the nucleosome array.

Could there be a functional role for localized, two-level nucleosome switching? Switching regions should be highly susceptible to any weak "signal" that affected the binding of nucleosomes to DNA, such as small changes in nucleosome concentration or affinity. Many organisms directly regulate nucleosome binding genome-wide, resulting in striking changes in the genome-wide average nucleosome density during cell differentiation. The most obvious purpose of a "nucleosome switch" would be the regulation of the access to DNA of regulatory proteins such as the abovementioned TF's. For example, particular TF binding sites could be blocked in cells with higher nucleosome densities but not in cells with lower nucleosome densities - or *vice-versa* -, which could play a role in cell differentiation. To test this hypothesis, we collected 278 published TF binding locations on chromosome II from the SGD database. We equilibrated the system at ~75% and ~90% average occupancies and calculated the absolute value of the difference in the probability of site occupancy. The mean (absolute) change was found to be 22.7%. When we calculated the mean absolute change in occupancy on the restricted set of the 278 TF binding locations, we found it to be 32.9%. To check for the statistical significance of this result, we randomly chose 250 different sets of 10 base pairs regions and calculated the average absolute change in site occupancy on changing the density. We repeated this procedure and generated a distribution of average absolute changes and found a standard deviation of 2% from the mean. This means that the statistical probability of randomly achieving a mean value of 32.9% is less than $10^{-7}$. It follows that at least *some* TF binding sites are strategically placed on segments of DNA that on average are more likely to reconfigure in response to changes in nucleosome concentration or affinity.

We conclude by noting some limitations of the model. Our basic premise was that the nucleosome array is in a state of thermal equilibrium. This could be questioned on grounds that the energy barriers between different binding sites - the variance of the $V_n$'s in effect - is of the order of tens of $k_BT$, which could lead to very long equilibration times in the dynamics of the system. In actuality, energy-consuming nucleosome positioning enzymes [9] may well be capable of "annealing" the array by generating an effective internal "noise" temperature that is in excess of the ambient temperature. Another important feature that was not included are spatial interactions between nucleosomes that are not nearest-neighbors. It is well known that the linear DNA/nucleosome array is coiled up into a "30 nm" fiber, which has an unknown, possibly zig-zag or solenoidal, spatial structure [2]. Spatial restrictions imposed on nucleosome positioning by the requirement that the array *can* be coiled up efficiently into a 30 nm fiber were not taken into account. Both questions are currently under investigation.

Acknowledgements: DJS thanks Dan Grilley for helpful discussions. RFB and DJS acknowledge support from the NSF under DMR Grant 04-04507. JW acknowledges support from NIGMS grants R01 GM054692 and R01 GM058617.

Figure Captions:

Figure 1: Correlation function $C(i) \propto \langle V_n V_{n+i} \rangle - \langle V_n \rangle^2$ for the onsite potentials $V_n$ of the nucleosomes of a section of chromosome II of budding yeast, shown as a function of the separation $i$, normalized so $C(0) = 1$. The fact that $C(i)$ does not vanish for large i is due to large-scale variations of the nucleosome concentration along the chromosome. The probability distribution of the $V_n$'s is approximately a Gaussian with a width of 9.3 $k_B T_0$.

Figure 2: Solid black line: Heat capacity of the nucleosome array of chromosome II at $\mu = 24$ $k_B T_0$ obtained from the iterative solution of Eqs. 1-3 as a function of the reduced temperature $T/T_0$. The heat capacity is expressed in units of $k_B L/a$, with L the system size and $a$ the hard core size. Above the heat capacity maximum, the site occupation probability (inset right, $T/T_0 = 8.0$) is approximately sinusoidal with a wavelength comparable to the hard-core size, as expected from linear response theory. Below the maximum, the occupation probability is, for most sites, either close to one or close to zero (inset left, $T/T_0 = 0.5$), consistent with nucleosome positioning. A significant minority of sites are unstable with intermediate occupancy probabilities. Grey solid line: percentage of sites that belong to an unstable sequence.

Figure 3: Site occupancy of an unstable section of chromosome II (bp 6500-9500) at $T/T_0 = 0.5$ for different values of the chemical potential. The mean occupancies are 88.8, 89.3, and 90%. For higher, respectively, lower chemical potentials, the occupation pattern is stable with, respectively, 11 and 10 nucleosomes. The unstable sequence for intermediate

chemical potential is the superposition of the two stable patterns.

Figure 1

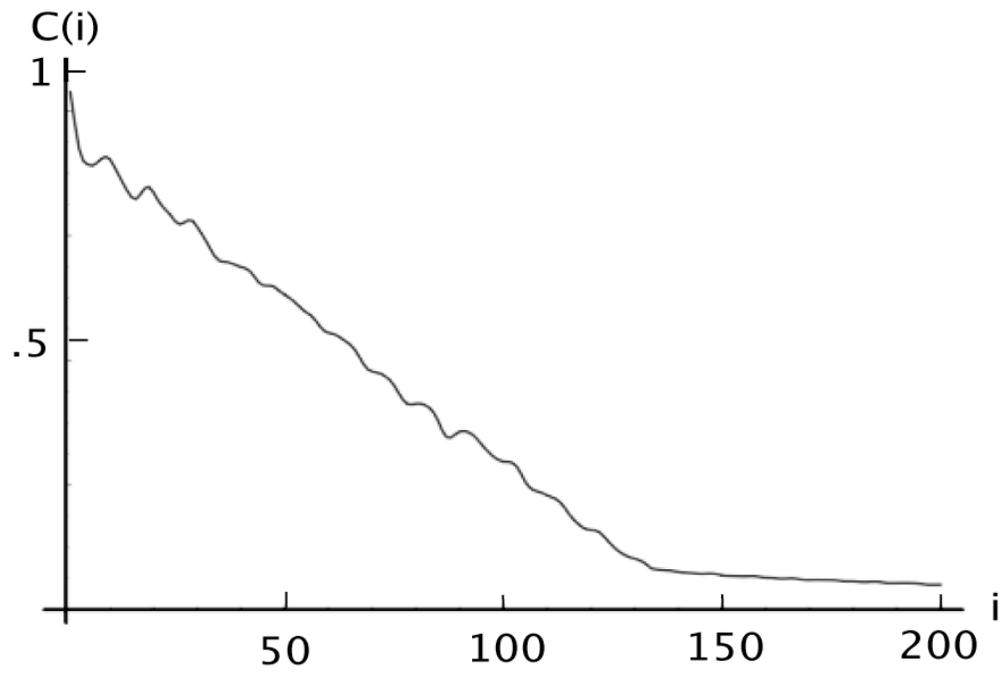

Figure 2

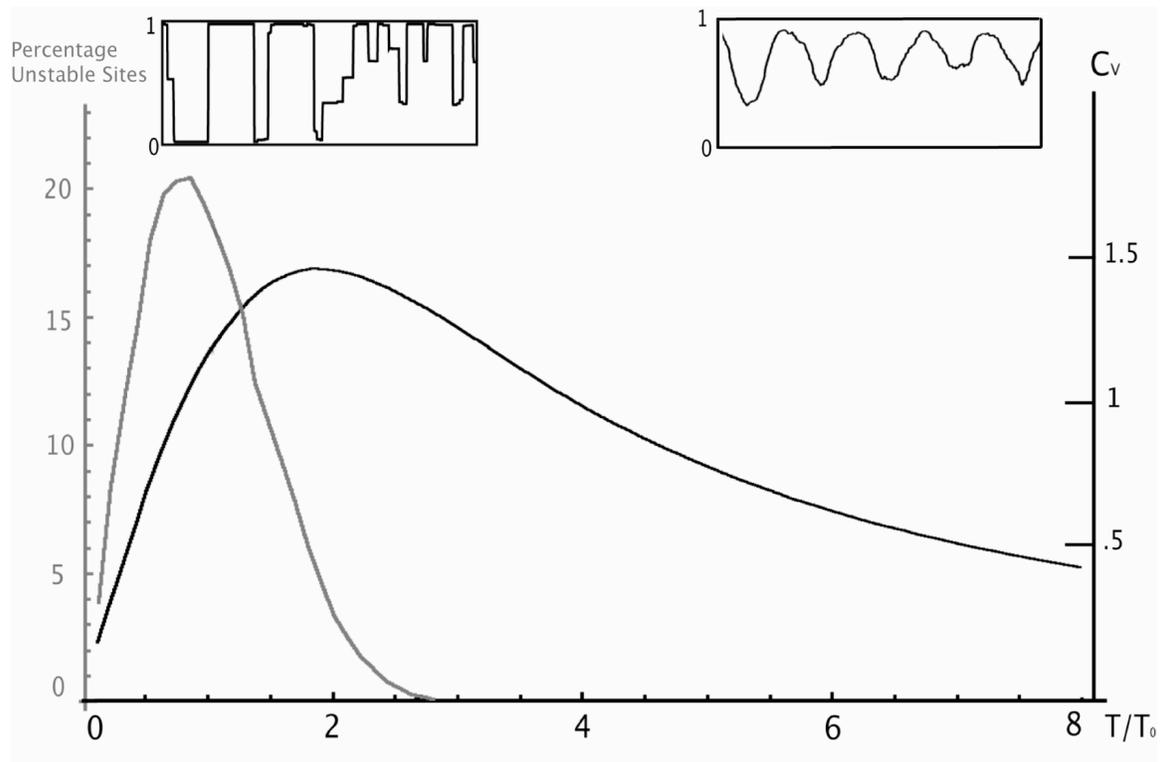

Figure 3

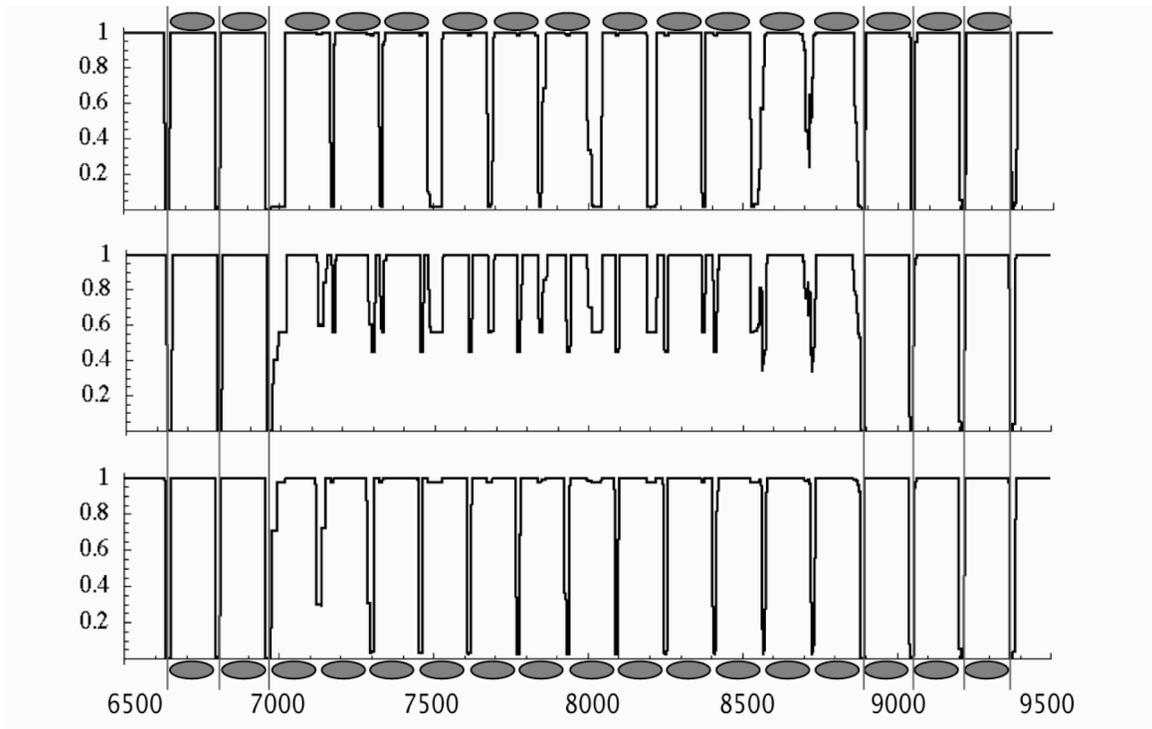